\documentclass[sigconf,screen]{acmart}

\acmSubmissionID{821}

\usepackage[inline]{enumitem}
\usepackage{xcolor,colortbl}
\usepackage{acronym}
\usepackage{bm}
\usepackage[skip=1mm]{caption}
\usepackage{natbib}
\usepackage{hyperref}
\hypersetup{
  colorlinks   = true,    
  urlcolor     = blue,    
  linkcolor    = blue,    
  citecolor    = red      
}
\definecolor{Gray}{gray}{0.85}
\definecolor{LightCyan}{rgb}{0.88,1,1}

\makeatletter
\newcommand\headingnodot{\def\@toclevel{4}%
  \@startsection{paragraph}{4}{\z@}%
  {-.2\baselineskip \@plus -2\p@ \@minus -.2\p@}%
  {-3.5\p@}%
  {\ACM@NRadjust{\bfseries}}}
\newcommand{\heading}[1]{\headingnodot{#1.}}
\makeatother

\AtBeginDocument{%
  }

\acrodef{DDRO}{direct document relevance optimization}
\acrodef{DSI}{differentiable search indexes}
\acrodef{GenIR}{generative information retrieval}
\acrodef{LM}{language model}
\acrodef{RLRF}{reinforcement learning from relevance feedback}
\acrodef{SFT}{supervised fine-tuning}

\setcopyright{rightsretained}
\copyrightyear{2025}
\acmYear{2025}
\setcopyright{cc}
\setcctype{by}
\acmConference[SIGIR '25]{Proceedings of the 48th International ACM SIGIR Conference on Research and Development in Information Retrieval}{July 13--18, 2025}{Padua, Italy}
\acmBooktitle{Proceedings of the 48th International ACM SIGIR Conference on Research and Development in Information Retrieval (SIGIR '25), July 13--18, 2025, Padua, Italy}
\acmDOI{10.1145/3726302.3730023}
\acmISBN{979-8-4007-1592-1/2025/07}

\author{Kidist Amde Mekonnen}
\orcid{0009-0006-1702-9514}
\affiliation{
  \institution{University of Amsterdam}
\city{Amsterdam}
  \country{The Netherlands}
}
\email{k.a.mekonnen@uva.nl}

\author{Yubao Tang}
\orcid{0009-0003-8010-3404}
\affiliation{
  \institution{University of Amsterdam}
\city{Amsterdam}
  \country{The Netherlands}
}
\email{y.tang3@uva.nl}

\author{Maarten de Rijke}
\orcid{0000-0002-1086-0202}
\affiliation{
 \institution{University of Amsterdam}
 \city{Amsterdam}
 \country{The Netherlands}
}
\email{m.derijke@uva.nl}

\renewcommand{\shortauthors}{Kidist Amde Mekonnen et al.}

\begin{document}
\title{Lightweight and Direct Document Relevance Optimization for Generative Information Retrieval}
\begin{abstract}
\Ac{GenIR} is a promising neural retrieval paradigm that formulates document retrieval as a document identifier (docid) generation task, allowing for end-to-end optimization toward a unified global retrieval objective. However, existing \ac{GenIR} models suffer from token-level misalignment, where models trained to predict the next token often fail to capture document-level relevance effectively. While reinforcement learning-based methods, such as \ac{RLRF}, aim to address this misalignment through reward modeling, they introduce significant complexity, requiring the optimization of an auxiliary reward function followed by reinforcement fine-tuning, which is computationally expensive and often unstable. To address these challenges, we propose \acfi{DDRO}, which aligns token-level docid generation with document-level relevance estimation through direct optimization via pairwise ranking, eliminating the need for explicit reward modeling and reinforcement learning. Experimental results on benchmark datasets, including MS MARCO document and Natural Questions, show that DDRO outperforms reinforcement learning-based methods, achieving a 7.4\% improvement in MRR@10 for MS MARCO and a 19.9\% improvement for Natural Questions. These findings highlight \ac{DDRO}'s potential to enhance retrieval effectiveness with a simplified optimization approach. By framing alignment as a direct optimization problem, \ac{DDRO} simplifies the ranking optimization pipeline of GenIR models while offering a viable alternative to reinforcement learning-based methods.
\end{abstract}



\maketitle

\acresetall

\section{Introduction}
\label{section:introduction}
Building on the success of transformer-based pre-trained language models, recent research has explored various neural retrieval approaches~\cite{yates-etal-2021-pretrained}: learned sparse retrieval \cite{formal2021splade-v2, Formal2021SPLADEVS}, dense retrieval \cite{Karpukhin2020DensePR, Xiong2020ApproximateNN, Zeng2022CurriculumLF, Zhan2021OptimizingDR}, and cross-encoders \cite{Nogueira2019PassageRW}. 
A new paradigm has recently been added to this palette, \acfi{GenIR}~\citep{Tay2022TransformerMA, Metzler2021RethinkingS}. 
This approach uses pre-trained encoder-decoder models as \ac{DSI}. 
It has inspired the development of several models \cite{NCI,Zhuang2022BridgingTG,mehta-etal-2023-dsi,bevilacqua2022autoregressive,Zhou2022UltronAU,Ren2023TOMEAT,Zeng2023ScalableAE,Zeng2024PlanningAI,semantic,tang2024listwise,NEURIPS2024_853e781c,tang2025generative,tang2024bootstrapped}


\heading{Generative information retrieval}
\ac{GenIR} models represent documents as sequences of \emph{unique document identifiers} (docids), generated autoregressively, where each token is conditioned on the query encoding and previously generated tokens. The generation process is controlled through (constrained) beam search \cite{mehta-etal-2023-dsi, Tay2022TransformerMA, NCI, Zeng2024PlanningAI, Zeng2023ScalableAE, Zhuang2022BridgingTG}. Docids could be predefined and remain static during training, making their careful design crucial for optimal retrieval performance \cite{recent}.
%
We classify docids into two categories based on their generation methodology and abstraction level. The first category, referred to as \emph{content-derived} docids, includes identifiers that are extracted directly from document elements such as titles \cite{DeCao2020AutoregressiveER, Chen2022GEREGE, Chen2022CorpusBrainPA, Lee2022NonparametricDF,10.1145/3626772.3661379, 10.1145/3589335.3641239,10.1007/978-3-031-56069-9_48,tang2023recent}, n-grams \cite{bevilacqua2022autoregressive, Chen2023AUG, Li2023LearningTR, li2023multiview, Wang2023NOVOLA}, URLs \cite{Ren2023TOMEAT, zhou-etal-2023-enhancing-generative, Zhou2022UltronAU, Ziems2023LargeLM}, and key terms \cite{Zhang2023TermSetsCB}. These docids preserve surface-level textual characteristics and are closely tied to the original document content. In contrast, the second category, termed \emph{computationally-generated} are derived using techniques like quantization \cite{chen2023continual, Rajput2023RecommenderSW, Zeng2023ScalableAE, Zhou2022UltronAU, Zeng2024PlanningAI} or hierarchical clustering algorithms \cite{mehta-etal-2023-dsi, Sun2023LearningTT, Tay2022TransformerMA, NCI} to encode deeper semantics by abstracting raw document content into conceptual features.

During training, \ac{GenIR} models learn to associate document text with corresponding docids, embedding semantic information directly into its parameters. During retrieval, docids are sequentially generated based on learned representations. By unifying indexing and retrieval within a transformer-based architecture, these models optimize both processes simultaneously~\cite{Metzler2021RethinkingS, Tay2022TransformerMA}. 

\heading{Challenges in \ac{GenIR}}
Despite recent advancements, \ac{GenIR} models face key limitations that hinder their effectiveness. These models typically rely on an auto-regressive loss function that optimizes the generation of individual docid tokens. However, this token-level optimization approach does not align with the broader goal of ranking tasks, which requires assessing the overall relevance of a document to the query. As a result, this misalignment often leads to suboptimal ranking performance.
\emph{To address these challenges, it is crucial to align token generation with document-level relevance estimation to ensure more accurate, well-rounded retrieval outcomes.}

\heading{Direct document-level relevance optimization}
Existing approaches aimed at aligning token-level docid generation with docu\-ment-level relevance estimation, such as \ac{RLRF}, use reinforcement learning to align predictions with relevance judgments through reward modeling~\citep{zhou-etal-2023-enhancing-generative}. However, \ac{RLRF} introduces significant complexity, requiring the optimization of an
auxiliary reward function followed by reinforcement fine-tuning, which is computationally expensive and often unstable. To address these challenges, we introduce a \acfi{DDRO} method that employs a pairwise ranking approach into \ac{GenIR} models to improve document retrieval performance. 

Our approach has two key phases. 
\begin{enumerate*}[label=(\roman*)]
\item First, we employ \ac{SFT} to train a \ac{LM} capable of generating docids that are most relevant to a given query. 
\item Next, we directly refine the model through pairwise ranking, where the model learns to differentiate between relevant and irrelevant docids for a specific query based on labeled data. This refinement ensures that the retrieval system ranks documents based on their relevance, aligning the model more closely with the query.
\end{enumerate*}
The \ac{SFT} phase serves as a pretraining step, aligning the model with initial relevance signals from training data and providing a strong foundation for the pairwise ranker to further fine-tune document ranking effectiveness.


Experiments conducted on the MS MARCO document ranking~\cite{bajaj2016ms} and Natural Questions (NQ)~\cite{Kwiatkowski2019NaturalQA} benchmarks demonstrate the effectiveness of \ac{DDRO} in improving retrieval accuracy, outperforming multiple baselines. 
Moreover, \ac{DDRO} maintains competitive performance with established baselines on broader metrics like R@10, demonstrating robustness across evaluation criteria. An ablation study further highlights the contributions of pairwise ranking optimization to the observed performance improvements.

\heading{Main contributions}
We introduce \acf{DDRO}, a pairwise ranking approach that aligns docid generation with document-level relevance judgments. This approach ensures that docids are generated not only based on token-level likelihood but also according to their relevance to the user's query. \ac{DDRO} unifies training objectives within a single framework, optimizing directly for document-level relevance. Experimental results demonstrate improvements in retrieval accuracy, highlighting the effectiveness of the proposed approach in enhancing generative retrieval models for relevance-based ranking.

\heading{Reproducibility}
To promote reproducibility in \ac{GenIR}, we open-source our codebase and make checkpoints publicly available.\footnote{\url{https://github.com/kidist-amde/DDRO-Direct-Document-Relevance-Optimization/tree/main}}

\section{Related Work}
\label{section:related-work}

\heading{Reward modeling}
Recent efforts in GenIR have sought to bridge the gap between token-level optimization and document-level relevance. GenRRL \cite{zhou-etal-2023-enhancing-generative} addresses this issue using \ac{RLRF} to align docid generation with query relevance. While effective, this approach requires a robust reward model training and reinforcement learning fine-tuning, both of which are resource-intensive and prone to instability. Developing a reliable reward model demands substantial labeled data, and reinforcement learning fine-tuning involves extensive hyperparameter tuning \cite{DPO}, contributing to training instability and scalability challenges for large-scale applications. In contrast, we propose \ac{DDRO}, a direct document-level relevance optimization method that eliminates the need for explicit reward model training and reinforcement learning fine-tuning, thereby reducing computational overhead and improving optimization efficiency.


\heading {Dense-generative integration}
Ranking-oriented generative retrieval (ROGER)~\citep{ROGER} combines dense and generative retrieval by using dense retrievers as intermediaries to provide relevance signals, bridging the gap between document ranking and docid generation. ROGER employs knowledge distillation from dense retrievers to enhance the generative model's ranking capabilities, combining the strong relative ranking signals of dense retrieval with the flexibility of generative models. However, it relies on external dense retrievers and does not directly optimize for document-level relevance within the generative model’s training objectives. In contrast, DDRO eliminates this dependency by incorporating pairwise ranking directly into the generative model’s optimization pipeline, ensuring alignment with document-level relevance.

\heading{Learning to rank in generative retrieval models}
Similarly, LTRGR~\citep{Li2023LearningTR} incorporates a learning-to-rank (LTR) framework to address the gap between docid generation and document ranking. It introduces an additional training phase where the model is optimized using a margin-based ranking loss, eliminating the need for a separate ranking step during inference. However, LTRGR focuses on optimizing passage ranking during the second phase, treating docid generation as a step toward this goal rather than fully integrating document-level relevance into the generative process. Consequently, the challenge of embedding document-level relevance directly into docid generation remains unaddressed. In contrast, DDRO integrates pairwise ranking directly into the generative model’s optimization pipeline, ensuring docid generation inherently aligns with document-level relevance.

\heading{Our approach}
Building on the advancements of GenRRL, ROGER, and LTRGR, we propose a framework that combines \ac{SFT} for docid generation with pairwise ranking optimization to better align GenIR objectives with ranking goals. \ac{DDRO} addresses token-level misalignment by incorporating document-level relevance optimization into the training process, enhancing existing GenIR systems by enabling them to learn to rank more effectively.

\heading{State-of-the-art baselines}
SOTA baselines in GenIR, such as RIPOR~\cite{Zeng2023ScalableAE} and PAG~\cite{Zeng2024PlanningAI}, employ multi-stage optimization approaches. 
E.g., RIPOR refines relevance-based docids through iterative pre-training and fine-tuning, while PAG introduces a hybrid decoding strategy that combines simultaneous and sequential decoding to enhance ranking efficiency. Both methods achieve strong results on the MS MARCO passage ranking dataset. Our work simplifies optimization with a single-framework approach, offering an alternative to multi-stage methods. A direct comparison with these baselines is deferred to future work to assess how \ac{DDRO} can complement and extend these approaches while evaluating its scalability in large-scale retrieval tasks.

\section{Preliminaries and Motivations} 
\subsection{Generative Information Retrieval (GenIR)}
\ac{GenIR} models build on large pre-trained language models, such as T5~\cite{raffel-2020-exploring} and BART~\cite{lewis-etal-2020-bart}, and are designed to take a query string and generate a ranked list of document identifiers (docids) based on their generation probabilities, ordered in descending sequence. Each document \( d \) is assigned a unique identifier \( \text{docid} \allowbreak = \allowbreak (\text{docid}_{1}, \allowbreak \text{docid}_{2}, \allowbreak \dots, \allowbreak \text{docid}_{L}) \), where \( L \) is the length of the identifier, and the model processes the query \( q \) to autoregressively generate the corresponding docid using a scoring function defined as: 
\begin{equation}
score(\text{docid} \mid q) = \prod_{i=1}^{L} p_{\theta}(\text{docid}_{i} \mid \text{docid}_{1:i-1}, q),
\end{equation}
where \( p_{\theta} \) denotes the generative retrieval model parameterized by \( \theta \), and \( \text{docid}_{i} \) is the \( i \)-th token of the docid for document \( d \). The generation continues until a special end-of-sequence (EOS) token is decoded. 

Training is performed using a multi-task setup that combines indexing and fine-tuning, which yields better results than training these tasks separately~\cite{Tay2022TransformerMA}.  During indexing, the model memorizes the document collection and maps each document's text to its corresponding docid.
Fine-tuning then refines this mapping by optimizing query-to-docid associations. The model is optimized via the following loss $L_{DSI}^{\theta}$ with teacher forcing \cite{Williams1989ALA}:
\begin{align}\label{eq:DSI}
 \sum_{d_i \in D} \log P({docid}_{i} \mid {T5}_{\theta}(d_i)) + \sum_{q_j \in Q} \log P({docid}_{i} \mid {T5}_{\theta}(q_j)),
\end{align}
where \( D \) represents the document set, and \( Q \) denotes the query set. This loss function enables parameter updates during both indexing and fine-tuning, enhancing the model’s ability to generate the most relevant docid for a given query or document. The retrieval phase employs a (constrained) beam search algorithm to decode the most probable docid, with their generation probabilities determining the final ranking \cite{mehta-etal-2023-dsi, Tay2022TransformerMA, NCI, Zeng2023ScalableAE, Zeng2024PlanningAI, Zhuang2022BridgingTG}.

\begin{figure*}[t] \centering \includegraphics[width=1.0\linewidth]{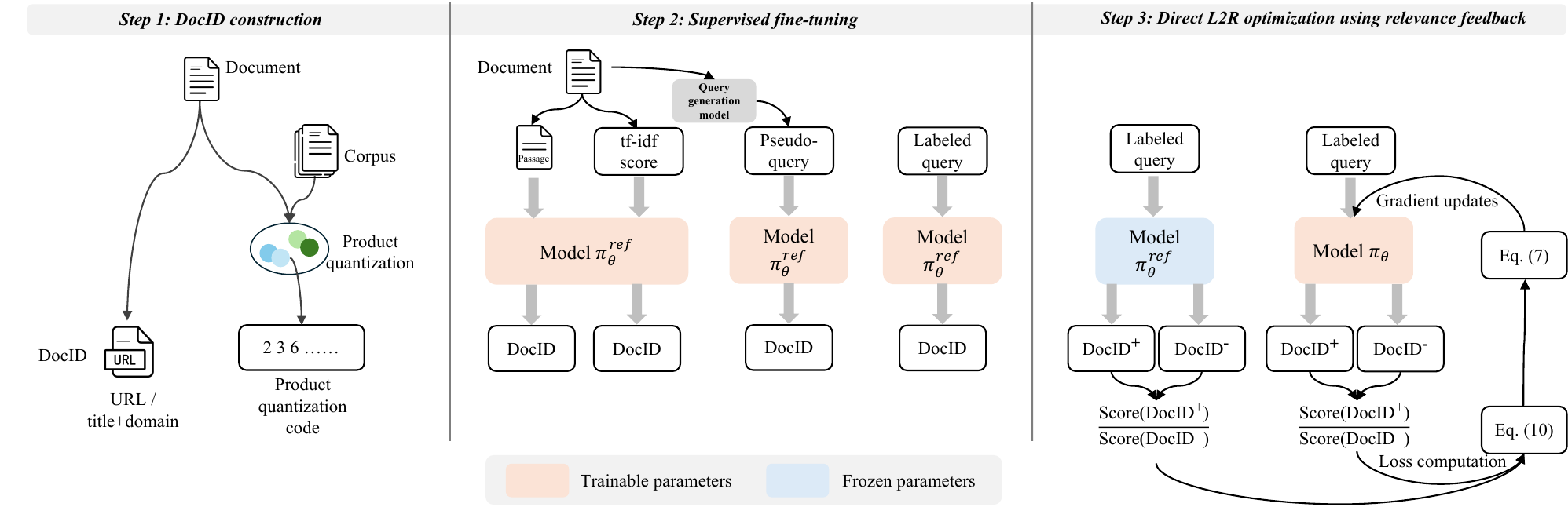} \caption{The proposed workflow comprises three key stages: (1) Construction of document identifiers (docids) including URL/title, domain, and product quantization codes; (2) Supervised fine-tuning of the retrieval model $\smash{\pi_\theta^\mathit{ref}}$ using diverse data pairs; and (3) Freezing the trained reference policy model $\smash{\pi_\theta^\mathit{ref}}$ and performing direct learning-to-rank (L2R) optimization on a policy model $\pi_\theta$.} \Description{The figure illustrates the workflow for document relevance optimization, including three steps: constructing DocIDs, supervised fine-tuning of the reference model, and direct learning-to-rank optimization.} \label{fig:workflow} \end{figure*}

\subsection{Learning to Rank (L2R)}
L2R aims at training models to rank documents based on their relevance to a given query~\cite{burges2010ranknet,10.1145/1273496.1273513listnet,10.1145/3269206.3271784lambdaloss}. L2R methods can be classified into point-wise, pair-wise, and list-wise approaches based on their learning objectives. 
\begin{enumerate*}[label=(\roman*)]
\item Point-wise 
methods~\cite{ibrahim2016comparing} frame ranking as a classification problem by scoring individual documents independently: $L_{\text{point}} = \sum_{i} \mathcal{L}(\hat{s}(d_i, q), s(d_i, q))$, where $\hat{s}(d_i, q)$ and $s(d_i, q)$ denote the predicted relevance score and ground truth relevance score, respectively. In GenIR, generated probabilities serve as relevance scores, aligning with this approach \cite{Tay2022TransformerMA}. And the retrieval term in Eq.~\ref{eq:DSI} belongs to this type.
\item Pair-wise approaches ~\cite{10.1145/3269206.3271784lambdaloss,dammak2017improving,burges2010ranknet,10.1145/1273496.1273513listnet} compare document pairs to determine relative preferences: 
$
L_{\text{pair}} = \sum_{(d_i, d_j)} \log \left( 1 + \exp \left( -\left( \hat{s}(d_i, q) - \hat{s}(d_j, q) \right) \right) \right)$, where $d_i$ and $d_j$ are used as pairs to compare. 
DDRO shares similarities with traditional pairwise L2R methods such as RankNet~\cite{RankNet} and LambdaRank~\cite{LambdaRank}, in that it optimizes a margin between relevant and non-relevant documents. It differs in that the ranking signal is used to supervise the generation of structured docid sequences via a generative decoder. Unlike typical L2R approaches that score documents retrieved by an external system, DDRO learns to produce docids directly, making it end-to-end generative. This integration of sequence modeling and pairwise supervision is a key distinction from prior L2R pipelines.
\item List-wise approaches ~\cite{xia2008listwiselistmle,lan2014positionplistmle}, optimize the entire ranked list:
$
L_{\text{list}} = \sum_{q} \mathcal{L}\left(\text{Softmax}(\hat{s}(q,\hat{\pi})), \text{Softmax}(s(q,\pi))\right)
$, 
where $\hat{\pi}$ and $\pi$ are the predicted and ground-truth lists, respectively. In GenIR, \citet{tang2024listwise} introduces a position-aware list-level objective to learn the relevance. 
As we focus on pair-wise approaches, comparisons with list-wise methods are left for future work.
\end{enumerate*}

A fundamental challenge in \ac{GenIR} stems from the inherent misalignment between the optimization objectives of autoregressive models and the overarching objectives of document ranking tasks. Training \ac{GenIR} models solely to generate docids can be treat as the point-wise approach, which is often insufficient for achieving effective ranking. Addressing this challenge necessitates the development of a robust framework that enables \ac{GenIR} models to directly learn to rank \cite{zhou-etal-2023-enhancing-generative, Li2023LearningTR,ROGER}.

\subsection{Reinforcement Learning from Relevance Feedback (RLRF)}
To address the aforementioned limitations, \citet{zhou-etal-2023-enhancing-generative} propose GenRRL, a generative retrieval model based on \ac{RLRF} to optimize generative models for alignment with document-level relevance. \ac{RLRF} optimizes rewards while ensuring alignment with human preferences using a KL divergence constraint \cite{zhou-etal-2023-enhancing-generative, Christiano2017DeepRL, bai2022constitutional, Stiennon2020LearningTS, Nakano2021WebGPTBQ, Li2023ReinforcementLW, Ouyang2022TrainingLM}. 
This method refines model predictions using a learned reward function, as formalized in prior research \cite{Jaques2016SequenceTC, Jaques2020HumancentricDT}: 
\begin{equation}
\max_{\pi_\theta} \mathbb{E}_{x \sim \mathcal{D}, y \sim \pi_\theta(y|x)} \left[ r_\phi(x, y) \right] - \beta \mathbb{D}_{\text{KL}}\left( \pi_\theta(y | x) \| \pi^{\text{ref}}(y | x) \right),
\end{equation}
where \( \beta \) is a parameter controlling deviation from the base reference policy \( \pi^{\text{ref}} \), which is typically the initial supervised fine-tuned model. This constraint prevents the model from straying too far from the data distribution used to train the reward function, preserving output diversity and avoiding overfitting to high-reward responses. Since language generation is non-differentiable, reinforcement learning (RL) techniques are widely used. A widely recognized approach \cite{Ziegler2019FineTuningLM, Stiennon2020LearningTS, Ouyang2022TrainingLM} optimizes the reward function using Proximal Policy Optimization (PPO) \cite{Schulman2017ProximalPO}, and defines the reward function as:
\begin{equation}
r(x, y) = r_\phi(x, y) - \beta (\log \pi_\theta(y | x) - \log \pi^{\text{ref}}(y | x)).
\end{equation} 
GenRRL \cite{zhou-etal-2023-enhancing-generative} trains a reward model using relevance-annotated data derived from BM25 \cite{Robertson2009ThePR}, DPR \cite{Karpukhin2020DensePR}, and LLaMA-13b \cite{Touvron2023LLaMAOA}. This reward model guides reinforcement learning to optimize the language model's policy for generating high-reward outputs, with a KL divergence constraint ensuring alignment with the original supervised fine-tuned model or the reference policy \( \pi^{\text{ref}} \). The optimization process involves supervised fine-tuning using negative log-likelihood, pairwise ranking loss for the reward model, and reinforcement learning techniques incorporating pointwise, pairwise, and listwise approaches to enhance ranking performance. While effective, this approach introduces considerable complexity, requiring the training of multiple models and sampling from the policy during training, which substantially increases computational costs.


Inspired by the work of~\citet{DPO}, we propose \acfi{DDRO}, a streamlined approach designed to enhance the ranking capabilities of \ac{GenIR} models. \ac{DDRO} directly optimizes GenIR models to learn document-level ranking without relying on explicit reward modeling or reinforcement learning.
While our optimization is inspired by the DPO framework~\cite{DPO}, its adaptation to GenIR is non-trivial. Unlike preference alignment for open-ended generation, our task involves optimizing structured docid generation under beam decoding constraints. Additionally, our method differs in both architecture (encoder-decoder vs. decoder-only) and objective (document ranking vs. preference alignment), requiring novel integration into GenIR pipelines.
To the best of our knowledge, DDRO is the first method to apply preference-style optimization directly to generative document retrieval by extending DPO-style training to constrained generative settings, leveraging pairwise query-docid relevance supervision and constrained decoding.
\section{Method}

We provide a comprehensive explanation of the proposed \ac{DDRO} method. As depicted in Figure (\ref{fig:workflow}), the method initiates with the generation of two categories of docids, designed to encapsulate diverse semantic and contextual features of the documents (Section \ref{sec:DocID-construction}). Subsequently, the retrieval model is trained through a combination of self-supervised and supervised learning techniques (Section \ref{sec:SFT}). Finally, direct learning-to-rank (L2R) optimization is applied, using relevance feedback to refine the model's ranking quality and align its outputs with document-level relevance (Section \ref{sec:Direct-L2R-optimization}).

\subsection{Docid Construction} \label{sec:DocID-construction}
The methodology for constructing docids in this work is grounded in established frameworks~\cite{Zhou2022UltronAU, zhou-etal-2023-enhancing-generative, Zeng2024PlanningAI, Zeng2023ScalableAE}. This approach uses keyword-based identifiers to effectively encapsulate the semantic and contextual information of the documents.

\noindent \textbf{URL and Title (TU).} 
 Titles in web search results are typically crafted to be descriptive, closely aligning with user search intent, while URLs often contain structured tokens, such as keywords or domains, that are highly indicative of relevance for web-based queries~\cite{Zhou2022UltronAU}. The structure of the URL reverses to prioritize semantically meaningful segments. 
 When a URL lacks descriptive content (e.g., uses numeric IDs or generic paths), we fall back to a combination of the document’s title and domain name as an alternative identifier. Formally, this docid variant is defined as:
\begin{equation}
\mbox{}\hspace*{-2mm}
docid_{\text{TU}} = 
\begin{cases}
\text{reverse(URL)}, &\text{if the URL is semantically rich}, \\ 
\text{title} + \text{domain}, &\text{otherwise}.
\end{cases}
\end{equation}

\heading{Product quantization codes (PQ)}
Building on prior work~\cite{chen2023continual, Rajput2023RecommenderSW, Zeng2023ScalableAE, Zhou2022UltronAU}, we adopt product quantization (PQ) to reduce the dimensionality of document representations while maintaining their semantic integrity. PQ compresses document vectors into latent semantic tokens by employing K-means clustering to partition the latent vector space into clusters. Each document is then represented by the corresponding cluster center, forming a compact identifier that preserves the document’s core semantic features. The resulting docid is defined as:
\begin{equation}
docid_{PQ} = PQ(\text{Encoder}(d)),
\end{equation}
where the encoder is based on a pre-trained T5 model  \cite{Ni2021SentenceT5SS}. The clustering process generates  $\bm{k}$  cluster centers across  $\bm{n}$ groups, expanding the vocabulary by $\bm{n \times k}$  new tokens. This approach produces a semantically rich and efficient representation of each document.

\subsection{Supervised Fine-tuning}\label{sec:SFT}
Supervised fine-tuning (SFT) enhances the retrieval capabilities of pre-trained language models by aligning them with task-specific data~\cite{Stiennon2020LearningTS, Ouyang2022TrainingLM, bai2022training}. 
Based on the two basic operations of DSI \cite{Tay2022TransformerMA}, i.e., indexing and retrieval tasks, diverse data pairs are curated and optimized using a teacher forcing policy~\cite{Williams1989ALA} to achieve alignment with the ground truth.

\begin{figure*}[!t]
\begin{minipage}{\textwidth}
\begin{equation}
\begin{split}
\nabla_\theta \mathcal{L}_{\mathrm{DDRO}}\left(\pi_\theta ; \pi^{\text{ref}}\right) =
-\beta \mathbb{E}_{\left(q, docid^+, docid^-\right) \sim \mathcal{D}}
& [\underbrace{\sigma\left(\hat{r}_\theta\left(q, docid^-\right)-\hat{r}_\theta\left(q, docid^+\right)\right)}_{\text{higher weight when reward estimate is wrong}} \times {} \\
&
\mbox{}\hspace*{5mm}
[\underbrace{\nabla_\theta \log \pi\left(docid^+ \mid q\right)}_{\text {increase likelihood of } docid^+} - \underbrace{\nabla_\theta \log \pi\left(docid^- \mid q\right)}_{\text {decrease likelihood of } docid^-}]],
\end{split}
\tag{7}
\label{eq:gradient-l2r-DDRO}
\end{equation}
\end{minipage}
\caption{Gradient for direct learning-to-rank optimization using relevance feedback.}
\Description{Mathematical equation describing the gradient update rule for direct learning-to-rank optimization using relevance feedback.}
\end{figure*}

\heading{Indexing task}
To memorize the corpus, the indexing task learns associations between documents and docids, making the document input format a crucial factor. Inspired by the indexing strategy proposed in \cite{Zhou2022UltronAU}, we generate self-supervised learning signals directly from the document corpus.

Text segments are mapped to their corresponding docids, enabling the model to link document passages with their broader context ~\cite{Callan1994PassagelevelEI, Zhou2022UltronAU}. Each document is divided into fixed-size passages, paired with the document's docid to create passage-to-docid pairs. For a document containing $N$ terms, $\{w_1, w_2, \ldots, w_N\}$, multiple passage-to-docid pairs are generated as follows:
\begin{equation}
\text{passage}: \{w_i, w_{i+1}, \ldots, w_{i+m-1}\} \rightarrow docid,
\end{equation}
where $i$ is the starting term of a passage, and 
$m$ is the fixed passage length.
To emphasize a document's core semantic content, terms are prioritized by their $tf$-$idf$ scores \cite{Robertson2009ThePR}, with a subset of high-scoring terms selected to form a compressed representation, which is then mapped to the document's docid:
\begin{equation}
\text{terms}: \{w_a, w_b, w_c\} \rightarrow docid,
\end{equation}
where $w_a, w_b, w_c$ are key terms selected based on their $tf$-$idf$ scores.

\heading{Retrieval task}
During retrieval, a pre-trained language model is fine-tuned with supervised query-docid pairs to learn semantic mappings between queries and their corresponding docids.  A key challenge in this process is the scarcity of labeled click data, which limits the ability to establish effective query-to-docid associations. To mitigate this, pseudo-queries are generated directly from the document corpus \cite{Zhuang2022BridgingTG, NCI}. Specifically, the docTTTTTquery~\cite{Cheriton2019FromDT} model is fine-tuned using supervised click data from the MS MARCO document and NQ datasets. For each document, an initial passage serves as input, and the model produces $k$ predicted queries using beam search, denoted as $Q = \{q_1, \ldots, q_k\}$. 

These diverse datasets, including passage-to-docid pairs, supervised query-docid pairs (derived from real-world relevance judgments), and synthetic pseudo-queries, are collectively used to train the $\bm{\pi}_{\theta}^{\text{ref}}$ model. Using these comprehensive and diverse training data, the SFT model acquires a robust understanding of query-to-document mappings and learns to generate relevant docids by optimizing token-level generation probabilities for a given query.

\heading{Training objective}
The model is trained with a sequence-to-sequence objective, aiming to maximize the likelihood of the target sequence through teacher forcing \cite{Williams1989ALA}. Given an input sequence $s$, which can be any of the document formats or queries described above, the objective is defined as:
\begin{equation}
L_{\text{SFT}}^{\theta} = \arg\max_{\theta} \log P(docid^* \mid s, \pi_{\theta}^{\text{ref}}(docid)),
\end{equation}
where $docid^*$ represents the ground truth sequence, $\pi_{\theta}^{\text{ref}}(docid)$ denotes the sequence generated by the \ac{SFT} model, and $P(docid^* \mid s, \pi_{\theta}^{\text{ref}}(docid))$ corresponds to the conditional probability of the ground truth given the input sequence and the model's generated sequence.

\subsection{Direct L2R Optimization Using Relevance Feedback}\label{sec:Direct-L2R-optimization}
\begin{figure}[t]
    \centering
    \includegraphics[width=1.0\linewidth]{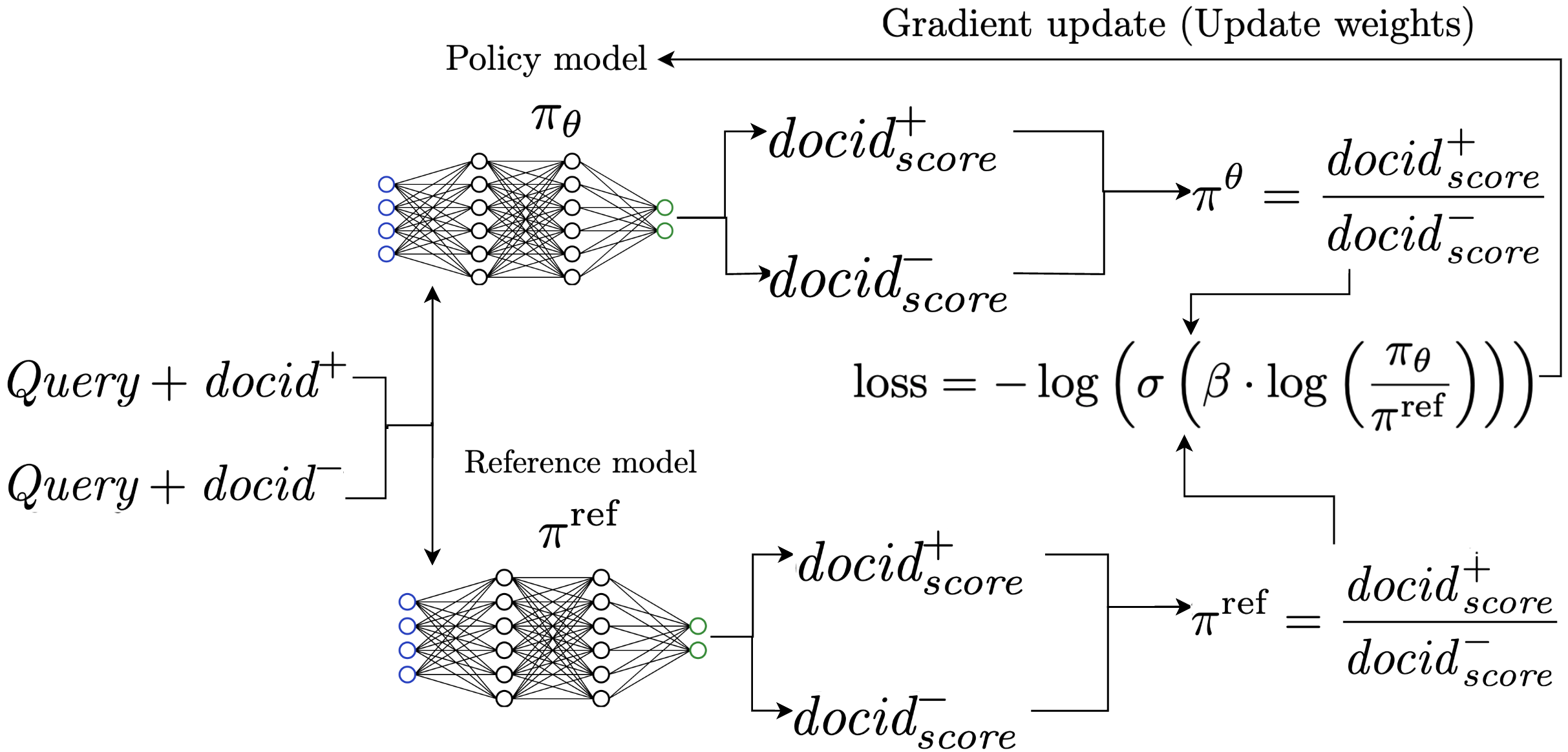} 
   \caption{Architecture of the \ac{DDRO} model, which fine-tunes the retrieval model through direct learning-to-rank (L2R) optimization using relevance feedback. Unlike GenRRL~\cite{zhou-etal-2023-enhancing-generative}, DDRO directly optimizes with relevance judgment data, avoiding reinforcement learning, explicit reward modeling, and extensive hyperparameter tuning. \textit{For clarity}, the model $\smash{\pi_\theta^\mathit{ref}}$ from the SFT phase is referred to as $\smash{\pi^\mathit{ref}}$, with its parameters frozen during this phase. }
    \Description{The architecture of the DDRO model highlights the fine-tuning process for the language model through Direct Document Relevance Optimization. It avoids reinforcement learning, reward modeling, and extensive hyperparameter tuning. The reference model $\pi_\theta^{ref}$ is shown with frozen parameters.}
    \label{fig:ddro_architecture}
\end{figure}

DDRO simplifies the complexities associated with reward modeling and reinforcement learning used in \ac{RLRF} approaches. Instead, it directly optimizes the likelihood that relevant docids ($docid^+$) are assigned higher scores over non-relevant ones ($docid^-$) for a given query, as illustrated in Figure~(\ref{fig:ddro_architecture}). The corresponding optimization objective is formulated as:
\begin{equation}
\mbox{}\hspace*{-2mm}
\begin{split}
&\mathcal{L}_{\text{DDRO}}(\pi_{\theta};\pi^{\text{ref}}) = {} -\mathbb{E}_{(q, docid^+, docid^-) \sim D} \\
&\left[ 
    \log \sigma \left( \beta \log \frac{\pi_{\theta}(docid^+ \mid q)}{\pi^{\text{ref}}(docid^+ \mid q)}
    - \beta \log \frac{\pi_{\theta}(docid^- \mid q)}{\pi^{\text{ref}}(docid^- \mid q)} \right) 
    \right],
\end{split}    
\hspace*{-2mm}\mbox{}
\end{equation}
where  \( \pi_\theta(docid | q) \) is the policy that is being optimized, while \( \pi^{\text{ref}}(docid | q) \) is the reference policy, typically the fine-tuned model (\ac{SFT}). This formulation ensures that the optimized model remains close to the reference policy while improving relevance-based ranking. The DDRO update guides the model toward producing outputs better aligned with relevance by utilizing pairwise comparisons, offering a simplified alternative to RL-based approaches. The gradient w.r.t.\ the model parameters $\theta$ is defined as in Eq.~\ref{eq:gradient-l2r-DDRO}, 
where
\begin{equation}
\hat{r}_\theta(q, docid) = \beta \log \frac{\pi_\theta(docid \mid q)}{\pi^{\text{ref}}(docid \mid q)}
\end{equation}
is the reward implicitly defined by the model \( \pi_\theta \) and the reference model \( \pi^{\text{ref}} \). The examples are weighted according to how much the implicit reward model \( \hat{r}_\theta \) overestimates the ranking of the non-relevant docid compared to the relevant docid. This weighting is scaled by \( \beta \), reflecting the degree of misjudgment while considering the strength of the KL divergence constraint. The sigmoid function \( \sigma(\cdot) \) ensures smooth optimization, and the model parameters \( \theta \) are adjusted to increase the likelihood of the relevant docids over non-relevant ones. This reparameterization streamlines the training process by eliminating the need for an explicit reward model and iterative fine-tuning, providing a more stable and efficient framework for aligning the retrieval model with document-level relevance objectives.

\section{Experimental Settings}
\subsection{Datasets and Evaluation Metrics}
\heading{Datasets}
We conduct our experiments using two widely recognized benchmarks: the 
\textbf{MS MARCO Document Ranking} dataset\footnote{\url{https://microsoft.github.io/msmarco/Datasets.html\#document-ranking-dataset}} \cite{bajaj2016ms} 
and the \textbf{Natural Questions (NQ)} dataset\footnote{\url{https://ai.google.com/research/NaturalQuestions/download}} \cite{Kwiatkowski2019NaturalQA}. The MS MARCO document ranking dataset is widely used for document ranking tasks and contains a large collection of queries and web pages. Following prior work \cite{zhou-etal-2023-enhancing-generative, Zhou2022DynamicRetrieverAP, Zhou2022UltronAU, Wang2023NOVOLA}, 
we use a subset with 320k documents and 360k query-document pairs for training. The NQ dataset, introduced by Google, is a widely used benchmark in question-answering research. In this study, we use the NQ320k version, which includes 320k query-document pairs sourced from Wikipedia, with queries formulated in natural language. To ensure reliable evaluation and improve performance \cite{lee-etal-2022-deduplicating}, we deduplicate documents by title and utilize the predefined training and validation splits.

\heading{Evaluation metrics}
Following \cite{zhou-etal-2023-enhancing-generative, Zhou2022DynamicRetrieverAP, Zhou2022UltronAU, Wang2023NOVOLA, ROGER}, we assess model performance using standard document retrieval metrics: Recall (R@1/5/ 10) and Mean Reciprocal Rank (MRR@10). Statistical significance is determined using paired t-tests with a threshold of \( p < 0.05 \).

\subsection{Baselines}
We evaluate our approach against three types of baseline: term-based retrieval, dense retrieval, and generative retrieval.
 
\heading{Term-based retrieval}  
\begin{enumerate*}[label=(\roman*)]  
\item \textbf{BM25}~\cite{bm25}, a probabilistic retrieval model commonly used as a standard baseline, implemented using Pyserini.\footnote{\url{https://github.com/castorini/pyserini}}
\item \textbf{DocT5Query}~\cite{Cheriton2019FromDT}, which generates synthetic queries from documents using the T5 model~\cite{raffel-2020-exploring}, appending them to the original document text.  
\end{enumerate*}  

\heading{Dense retrieval}  
\begin{enumerate*}[label=(\roman*)]  
\item \textbf{DPR}~\cite{Karpukhin2020DensePR}, which utilizes a BERT-based dual encoder to produce dense embeddings for queries and documents. PseudoQ~\cite{tang2021improving} improves DPR by generating pseudo-queries using K-means clustering over document embeddings.  
\item \textbf{ANCE}~\cite{Xiong2020ApproximateNN}, a RoBERTa-based dual encoder that incorporates hard negatives retrieved from an asynchronously updated approximate nearest neighbor (ANN) index.  
\item \textbf{RepBERT}~\cite{zhan2020repbert}, a BERT-based model that generates fixed-length contextualized embeddings, with query-document relevance computed via inner product similarity.  
\item \textbf{Sentence-T5}~\cite{Ni2021SentenceT5SS}, which applies a T5-based architecture to generate sentence embeddings using encoder-only and encoder-decoder models with contrastive learning.  
\end{enumerate*} 

\heading{Generative retrieval}  
\begin{enumerate*}[label=(\roman*)] 
\item \textbf{DSI}~\cite{Tay2022TransformerMA}, which represents docids using hierarchical k-means cluster IDs and trains with the DSI-Num objective.  
\item \textbf{DSI-QG}~\cite{Zhuang2022BridgingTG}, which augments training data with synthetic queries generated using a query generation model~\cite{Cheriton2019FromDT} and represents documents with arbitrary unique numerical docids.
\item \textbf{NCI}~\cite{NCI}, which assigns semantically structured numeric docids paired with pseudo-queries.  
\item \textbf{SEAL}~\cite{DeCao2020AutoregressiveER}, which retrieves docids represented as arbitrary n-grams extracted from document text using an FM-index.  
\item \textbf{Ultron}~\cite{Zhou2022UltronAU}, which employs keyword and semantic-based docids, using a three-stage training approach: general pre-training, search-oriented pre-training, and supervised fine-tuning.  
\item \textbf{ROGER}~\cite{ROGER}, which transfers document relevance knowledge from a dense retriever to a generative retriever via knowledge distillation.  
\item \textbf{MINDER}~\cite{li2023multiview}, which assigns multiple identifiers, including titles, n-grams, and synthetic queries, to documents and pairs them for indexing.
\item \textbf{LTRGR}~\cite{Li2023LearningTR}, which trains on pairwise relevance objectives using margin-based ranking loss for optimization.  
\item \textbf{GenRRL}~\cite{zhou-etal-2023-enhancing-generative}, which incorporates pointwise, pairwise, and listwise relevance optimization through reinforcement learning, using document summaries and URLs as docids.   
\end{enumerate*}  
We exclude document summaries as docids due to their dependence on external summarization models, such as LLaMA-13b used in GenRRL. These models introduce preprocessing overhead and variability in identifier quality. Instead, DDRO employs product quantization (PQ) to generate compact, structured docids, ensuring consistency and scalability. 

\heading{Note on result sourcing} Baseline results for methods such as GenRRL~\cite{zhou-etal-2023-enhancing-generative} and ROGER~\cite{ROGER} are taken from their original papers due to the unavailability of public code, ensuring consistency and avoiding potential reproducibility issues. Other results were reproduced using publicly available code and the dataset configurations described in this work.

\subsection{Implementation Details}
\textbf{SFT.} The SFT model is based on the T5-base pretrained model \cite{raffel-2020-exploring}, trained with a learning rate of \texttt{1e-3} and a batch size of $128$. All experiments involving various docid types were conducted on $8$ NVIDIA RTX A6000 GPUs.

\heading{Pseudo queries} The DocT5Query model \cite{Cheriton2019FromDT}, fine-tuned on the target dataset with document-query pairs, was used to generate $10$ pseudo-queries per document.

\heading{Contrastive data pair construction} Training triples were generated using stratified sampling for diversity. Positive samples were selected based on qrels relevance judgments, while negatives were drawn from the top 1000 BM25-retrieved documents, stratified into top (1–100), mid (101–500), and lower (501–1000) ranks. Negatives were randomly sampled in roughly equal proportions, with 8 per query for NQ and 16 for MS MARCO.

\heading{DDRO} The \ac{DDRO} model was initialized with the pre-trained autoregressive \ac{SFT} model (see Section \ref{sec:SFT}) and fine-tuned using the proposed direct learning-to-rank framework (see Section \ref{sec:Direct-L2R-optimization}). Training was performed using a modified Hugging Face TRL \texttt{DPOTrainer} \cite{vonwerra2022trl}, adapted for encoder-decoder models. A cosine learning rate scheduler with 1000 warm-up steps and early stopping was applied. The learning rate was set to \texttt{5e-6} for PQ-based docids and \texttt{1e-5} for URL-based docids, with a batch size of $64$ and a regularization parameter $\beta$ of $0.4$ to balance chosen and rejected responses. All experiments were conducted on a single NVIDIA A100 GPU.

\heading{Constrained beam search} During inference, constrained beam search generates valid docids by using a prefix tree~\cite{Tay2022TransformerMA} to enforce valid token sequences.

\section{Experimental Evaluation and Results}
Our evaluation of \ac{DDRO} focuses on the following questions:

\begin{enumerate}[label=\textbf{RQ\arabic*},nosep,leftmargin=*]
   \item How does \ac{DDRO} compare to RLRF-based methods, such as GenRRL, in terms of retrieval performance while avoiding the complexities of reward modeling and reinforcement learning?
   \item How does \ac{DDRO} perform relative to established baselines in terms of retrieval accuracy and ranking consistency on benchmark datasets?
   \item What is the impact of pairwise ranking optimization on the performance of \ac{DDRO}?
    \item How robust is \ac{DDRO} across datasets with varying characteristics?
     \item How does \ac{DDRO} balance relevance across the ranked list in generative retrieval models, and what impact does it have on overall ranking quality?

\end{enumerate}
\noindent

\subsection{Comparison with Reinforcement Learning-Based Methods}
To address RQ1, we compare \ac{DDRO} with GenRRL~\cite{zhou-etal-2023-enhancing-generative} on both datasets. 
Results are presented in Table~\ref{tab:genrrl_ddro_ms_marco} and Table~\ref{tab:genrrl_ddro_natural_questions}.
%

%
\begin{table}[t]
\centering
\caption{Performance comparison of GenRRL and DDRO on the MS MARCO document ranking (MS300K) dataset. The best results are in bold. Results for cited models are sourced from their original papers. Abbreviations: PQ -- Product Quantization; TU -- Title + URL; Sum -- document summary.}
\label{tab:genrrl_ddro_ms_marco}
\begin{tabular}{lcccc}
\toprule
\textbf{Model} & \textbf{R@1} & \textbf{R@5} & \textbf{R@10} & \textbf{MRR@10} \\
\midrule
GenRRL (TU)~\cite{zhou-etal-2023-enhancing-generative} & 33.01 & 63.62 & {74.91} & 45.93 \\
GenRRL (Sum) ~\cite{zhou-etal-2023-enhancing-generative} & 33.23 & 64.48 & \textbf{75.80} & 46.62 \\
DDRO (PQ) & 32.92 & {64.36} & 73.02 & 45.76 \\
DDRO (TU) & \textbf{38.24} & \textbf{66.46} & 74.01 & \textbf{50.07} \\
\bottomrule
\end{tabular}
\end{table}

We observe the following:
\begin{enumerate*}[label=(\roman*)]
\item On MS MARCO, \ac{DDRO} (TU) achieves the highest scores in early precision-focused metrics. Specifically, it outperforms GenRRL(Sum) by 15.06\% in R@1 and 7.4\% in MRR@10, highlighting its effectiveness in ranking the most relevant document at the top. While GenRRL (Sum) performs better in the R@10 metric, \ac{DDRO} (TU) achieves comparable results with a simplified optimization process, avoiding the need for a reward model and reinforcement learning.
\item On NQ, \ac{DDRO} (PQ) outperforms GenRRL (Sum) by 34.69\% in R@1 and 19.87\% in MRR@10, further validating its effectiveness in retrieving relevant documents within the top ranks. GenRRL variants achieve higher R@10 scores,  likely benefiting from multi-signal learning and listwise optimization strategies, potentially improving broader document retrieval. Summary-based docids may aid performance on knowledge-intensive queries. 

\item DDRO (TU) and DDRO (PQ) show different trends across datasets: TU performs better on MS300K, while PQ excels on NQ. This disparity likely stems from higher document quality in NQ, where the rich semantic content allows the generated PQs to convey more meaningful information. In contrast,  MS300K, which is derived from web search logs, contains noisy content such as ads, resulting in lower-quality PQs. Consequently, TU, which prioritizes keyword-based features, effectively captures the core content of MS300K, where the shorter, keyword-focused queries align well with surface-level signals such as titles and URLs.
\end{enumerate*}

\begin{table}[t]
\centering
\caption{Performance comparison of GenRRL and DDRO on the NQ320K dataset. The best results are in bold. Results for cited models are sourced from their original papers. 
Abbreviations: PQ -- Product Quantization; TU -- Title + URL; Sum -- document summary.}
\label{tab:genrrl_ddro_natural_questions}
\begin{tabular}{lcccc}
\toprule
\textbf{Model} & \textbf{R@1} & \textbf{R@5} & \textbf{R@10} & \textbf{MRR@10} \\
\midrule
GenRRL (TU)~\cite{zhou-etal-2023-enhancing-generative}   & 35.79 & {56.49} & {70.96} & 45.73 \\
GenRRL (Sum) ~\cite{zhou-etal-2023-enhancing-generative}  & 36.32&57.42 & \textbf{71.49} & 46.31\\
DDRO (TU) & {40.86} & 53.12 & 55.98 & 45.99 \\
DDRO (PQ) & \textbf{48.92} & \textbf{64.10} & {67.31} & \textbf{55.51} \\
\bottomrule
\end{tabular}
\end{table}

\subsection{Comparison with Established Baselines}

To address RQ2, Table~\ref{tab:msmarco_results} presents a comprehensive comparison of \ac{DDRO} with baselines on the MS300K dataset. We can observe the followings:
\begin{enumerate*}[label=(\roman*)]
\item The performance of dense retrieval baselines is generally better than that of sparse retrieval baselines, likely because the former uses dense vectors to capture richer semantic information, which is consistent with earlier findings \cite{lu-etal-2021-lessismore,10.1145/3477495.3531772costa}.

\item The best-performing dense retrieval baseline, ANCE, outperforms others such as SEAL, NCI, and DSI-QG. This could be due to the fact that these generative retrieval baselines rely solely on maximum likelihood estimation (MLE) to learn relevance, which may not fully capture the relevance patterns. However, ANCE lags behind models like Ultron, ROGER, and LTGR, which employ more advanced optimization strategies. This highlights the need for fine-grained relevance modeling to enhance generative retrieval ranking performance.

\item Our proposed DDRO (TU) outperforms these generative retrieval baselines, achieving 15.63\% and 8.03\% higher R@1 and MRR@10, respectively, compared to the best-performing baseline, ROGER-Ultron. These results demonstrate the effectiveness of our document-level relevance optimization approach. While ROGER-Ultron achieves the highest R@10, \ac{DDRO} (TU) delivers comparable performance within a more efficient and lightweight framework. 
\end{enumerate*}

\begin{table}[t]
\centering
\caption{Comparison of retrieval model performance on the MS MARCO document ranking (MS300K) dataset. The best results are highlighted in bold. Statistical significance is assessed using a paired t-test with a $p < 0.05$ threshold, where improvements are marked with the dagger symbol ($^\dagger$) to indicate statistical significance.  The second-best values are underlined. Results for cited models are sourced from their original papers. Abbreviations: SI -- Semantic ID; PQ -- Product Quantization; NG -- N-grams; TU -- Title + URL.}
\label{tab:msmarco_results}
\setlength{\tabcolsep}{1.2mm}
\begin{tabular}{lcccc}
\toprule
\textbf{Model}  & \textbf{R@1} & \textbf{R@5} & \textbf{R@10} & \textbf{MRR@10} \\
\midrule
\multicolumn{5}{l}{\emph{Term-based retrieval}} \\
BM25  & 18.94 & 42.82 & 55.07 & 29.24 \\
DocT5Query & 23.27 & 49.38 & 63.61 & 34.81 \\
\midrule
\multicolumn{5}{l}{\emph{Dense retrieval}} \\
DPR & 29.08 & 62.75 & 73.13 & 43.41 \\
ANCE & 29.65 & 63.43 & \underline{74.28} & 44.09 \\
RepBERT & 25.25 & 58.41 & 69.18 & 38.48 \\
Sentence-T5 & 27.27 & 58.91 & 72.15 & 40.69 \\
\midrule
\multicolumn{5}{l}{\emph{Generative retrieval}} \\
DSI (SI) & 25.74 & 43.58 & 53.84 & 33.92 \\
DSI-QG (SI) & 28.82 & 50.74 & 62.26 & 38.45 \\
NCI (SI) & 29.54 & 57.99 & 67.28 & 40.46 \\
SEAL (NG) & 27.58 & 52.47 & 61.01 & 37.68 \\
Ultron (TU) & 29.82 & 60.39 & 68.31 & 42.53 \\
Ultron (PQ) & 31.55 & 63.98 & 73.14 & 45.35 \\
ROGER-NCI (SI)~\cite{ROGER} & 30.61 & 59.02 & 68.78 & 42.02 \\
ROGER-Ultron (TU)~\cite{ROGER} & 33.07 & 63.93 & \textbf{75.13} & 46.35 \\
{MINDER} (SI)  & 29.98 & 58.37 & 71.92 & 42.51 \\
{LTRGR} (SI)  & 32.69 & 64.37 & 72.43 & \underline{47.85} \\
\midrule
\multicolumn{5}{l}{\emph{Ours}} \\
{DDRO} (PQ) & 32.92 & 64.36 & 73.02 & 45.76 \\
{DDRO} (TU) & \textbf{38.24}\rlap{$^\dagger$} & \textbf{66.46}\rlap{$^\dagger$} & \underline{74.01} & \textbf{50.07}\rlap{$^\dagger$} \\
\bottomrule
\end{tabular}
\end{table}

\subsection{Ablation Study}  
To address RQ3, an ablation study was conducted to assess the impact of pairwise ranking optimization on \ac{DDRO} performance. From Table~\ref{tab:ablation-study}, we can find:
\begin{enumerate*}[label=(\roman*)]
\item On the MS MARCO dataset, removing pairwise ranking optimization significantly reduces intermediate and broader recall metrics (R@5 and R@10) for both \ac{DDRO} variants, highlighting its critical role in improving retrieval performance.
\item On the NQ dataset, excluding pairwise ranking optimization leads to consistent declines across all metrics, with a more pronounced impact on DDRO (PQ), particularly in early precision and broader recall. This underscores the importance of pairwise ranking in enhancing retrieval effectiveness at various ranking depths.
\end{enumerate*}

These findings confirm that pairwise ranking optimization effectively aligns model predictions with document relevance, contributing to improved performance across different ranking levels.

\begin{figure}[t]
\centering
\includegraphics[width=1\linewidth]{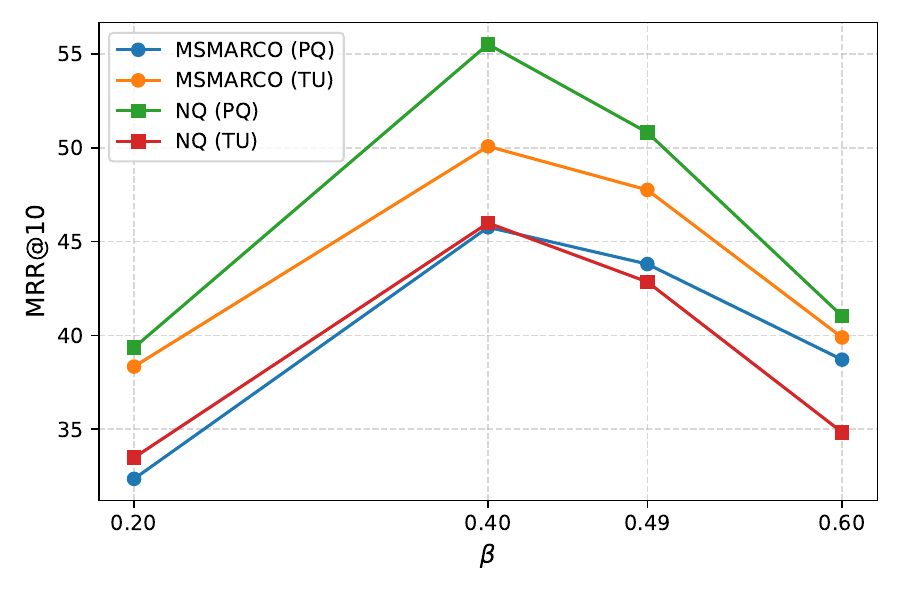}
\Description{A line plot showing MRR@10 across different beta values. Moderate beta (0.4) yields the highest performance on MS MARCO and Natural Questions.}
\caption{Effect of KL constraint strength ($\beta$) on DDRO performance. A moderate value ($\beta = 0.4$) yields the best MRR@10, while under- or over-regularization degrades performance.}
\label{fig:beta_sensitivity}
\end{figure}

\heading{Effect of KL Constraint Strength $\beta$}
We evaluate the impact of the $\beta$ parameter, which controls the KL divergence constraint between the DDRO policy $\smash{\pi_\theta}$ and the reference policy $\smash{\pi^\mathit{ref}}$, on retrieval performance.
Figure~\ref{fig:beta_sensitivity} shows results for different $\beta$ values on MS MARCO and Natural Questions (NQ). A moderate setting ($\beta = 0.4$) consistently yields the best MRR@10 across both datasets. Smaller values (e.g., $\beta = 0.2$) lead to under-regularization, resulting in unstable and suboptimal learning. In contrast, larger values (e.g., $\beta = 0.6$) impose excessive regularization, restricting the model’s ability to adapt, and thus degrading performance. These results highlight the sensitivity of DDRO to the KL constraint, suggesting the importance of tuning $\beta$ to balance learning flexibility and regularization.


\subsection{Robustness Analysis Across Datasets}
To address RQ4, additional experiments were conducted NQ dataset to evaluate the robustness of \ac{DDRO} across datasets with varying characteristics. The analysis focuses on two aspects: 
\begin{enumerate*}[label=(\roman*)]
\item comparing \ac{DDRO} retrieval performance with baseline models across different categories Table~\ref{tab:nq_results}, and 
\item examining the impact of various docid design choices on retrieval effectiveness.
\end{enumerate*}

\textit{\textbf{Comparison to baseline retrieval models.}} The performance comparison on the NQ320k across different  retrieval baselines is as follow: 
\begin{enumerate*}[label=(\roman*)]
\item {Term-based baselines}, such as BM25 and DocT5Query, show lower early ranking performance, but remain competitive at broader levels.
\item {Dense retrieval baselines}, including DPR and ANCE, improve early ranking metrics over term-based methods.
\item {Generative retrieval baselines}, such as ROGER-Ultron (TU) and LTRGR, perform well across all metrics. 
The proposed \ac{DDRO} achieves the highest overall performance. Specifically, DDRO (PQ) surpasses the best-performing baseline, ROGER-Ultron (TU), by 36.27\% in R@1 and 23.58\% in MRR@10.
\end{enumerate*}

\begin{table}[t]
\caption{Ablation study evaluating the impact of pairwise ranking optimization on \ac{DDRO} performance across the MS300K and NQ320K datasets. Statistical significance is assessed using a paired t-test with a significance threshold of \( p < 0.05 \). Statistically significant improvements (\( p < 0.05 \)) are marked with a dagger symbol ($^{\dagger}$), while non-significant improvements are underlined. Abbreviations: PQ -- Product Quantization; TU -- Title + URL.}

\label{tab:ablation-study}
\centering
\setlength{\tabcolsep}{1.2mm}  
\begin{tabular}{lcccc}
    \toprule
    \multicolumn{5}{c}{\textbf{MS MARCO doc}} \\
    \cmidrule{1-5}
    \textbf{Model} &  \textbf{R@1} & \textbf{R@5} & \textbf{R@10} & \textbf{MRR@10} \\
    \midrule
    {DDRO (PQ)} & \underline{32.92} & 64.36\rlap{$^{\dagger}$} & 73.02\rlap{$^{\dagger}$} & \underline{45.76} \\
    \hspace{3mm} w/o pairwise ranking & 32.18 & 62.62 & 71.29 & 44.79 \\
    {DDRO (TU)} & \underline{38.24} & 66.46\rlap{$^{\dagger}$} & 74.01\rlap{$^{\dagger}$} & \underline{50.07} \\
    \hspace{3mm} w/o pairwise ranking & 38.12 & 64.60 & 72.90 & 49.18 \\
    \midrule
    \multicolumn{5}{c}{\textbf{Natural Questions}} \\
    \cmidrule{1-5}
    \textbf{Model} &  \textbf{R@1} & \textbf{R@5} & \textbf{R@10} & \textbf{MRR@10} \\
    \midrule
    {DDRO (PQ)} & 48.92\rlap{$^{\dagger}$}  & 64.10\rlap{$^{\dagger}$}  & 67.31\rlap{$^{\dagger}$} & 55.51\rlap{$^{\dagger}$} \\
    \hspace{3mm} w/o pairwise ranking & 44.19 & 58.44 & 62.23 & 50.48 \\
    {DDRO (TU)} & 40.86\rlap{$^{\dagger}$} & 53.12\rlap{$^{\dagger}$} & 55.98\rlap{$^{\dagger}$} & 45.99\rlap{$^{\dagger}$} \\
    \hspace{3mm} w/o pairwise ranking & 39.58 & 50.50 & 53.53 & 44.32 \\
    \bottomrule
\end{tabular}
\end{table}

\begin{table}[t]
\centering
\caption{Comparison of retrieval model performance on the NQ320K dataset. The best-performing results are shown in bold. Statistical significance is determined using a paired t-test with a significance threshold of \( p < 0.05 \), with a dagger symbol ($^{\dagger}$) indicating statistical significance. Results for cited models are drawn from their respective original publications. Abbreviations used: SI -- Semantic ID; PQ -- Product Quantization; NG -- N-grams; TU -- Title + URL.}
\label{tab:nq_results}
\setlength{\tabcolsep}{1.2mm}
\begin{tabular}{lcccc}
\toprule
\textbf{Model}  & \textbf{R@1} & \textbf{R@5} & \textbf{R@10} & \textbf{MRR@10} \\
\midrule
\multicolumn{5}{l}{\emph{Term-based retrieval}} \\
BM25  & 14.06 & 36.91 & 47.93 & 23.60 \\
DocT5Query & 19.07 & 43.88 & 55.83 & 29.55 \\
\midrule
\multicolumn{5}{l}{\emph{Dense retrieval}} \\
DPR & 22.78 & 53.44 & 68.58 & 35.92 \\
ANCE & 24.54 & 54.21 & 69.08 & 36.88 \\
RepBERT & 22.57 & 52.20 & 65.65 & 35.13 \\
Sentence-T5 & 22.51 & 52.00 & 65.12 & 34.95 \\
\midrule
\multicolumn{5}{l}{\emph{Generative retrieval}} \\
DSI (SI) & 27.42 & 47.26 & 56.58 & 34.31 \\
DSI-QG (SI) & 30.17 & 53.20 & 66.37 & 38.85 \\
NCI (SI) & 32.69 & 55.82 & 69.20 & 42.84 \\
SEAL (NG) & 29.30 & 54.12 & 68.53 & 40.34 \\
Ultron (TU) & 33.78 & 54.20 & 67.05 & 42.51 \\
Ultron (PQ) & 25.64 & 53.09 & 65.75 & 37.12 \\
ROGER-NCI (SI)~\cite{ROGER} & 33.20 & 56.34 & 69.80 & 43.45 \\
ROGER-Ultron (TU)~\cite{ROGER} & 35.90 & 55.59 & \textbf{69.86} & 44.92 \\
{MINDER} (SI)  & 31.00 & 55.50 & 65.79 & 43.50 \\
{LTRGR} (SI)  & 32.80 & 56.20 & 68.74 & 44.80 \\
\midrule
\multicolumn{5}{l}{\emph{Ours}} \\
{DDRO} (TU) & 40.86 & 53.12 & 55.98 & 45.99 \\
{DDRO} (PQ) & \textbf{48.92}\rlap{$^\dagger$} & \textbf{64.10}\rlap{$^\dagger$} & 67.31 & \textbf{55.51}\rlap{$^\dagger$} \\
\bottomrule
\end{tabular}
\end{table}

\textit{\textbf{Impact of Dataset Characteristics and Docid Selection.}}  
The docid design is a critical factor influencing performance in GenIR, we further analyze the differences in performance among various designs within our proposed \ac{DDRO}.
Our analysis underscores the critical impact of docid design on retrieval performance across datasets. 
\ac{DDRO} (TU) excels on MS MARCO, where shorter, keyword-driven queries align well with title and URL-based docids that capture surface-level lexical features for efficient retrieval. 
In contrast, \ac{DDRO} (PQ) performs better on NQ, which features longer, complex queries requiring deeper semantic understanding. 
PQ-based docids effectively capture latent relationships, making them well-suited for NQ's informational queries. These findings suggest that aligning docid strategies with dataset-specific characteristics enhances retrieval effectiveness and model adaptability.

\subsection{Analysis of Relevance Distribution}
The relevance distribution could, to some extent, reflect the retrieval model's ability to recognize relevance. Therefore, we analyze the retrieval performance of DDRO at different top positions in the generated docid list to address RQ5.
We have the following observations:
\begin{enumerate*}[label=(\roman*)]
\item Tables~(\ref{tab:msmarco_results}) and~(\ref{tab:nq_results}) demonstrate \ac{DDRO}'s effectiveness across datasets, showcasing its adaptability. On MS MARCO, \ac{DDRO} (TU) achieves the highest early precision, with statistically significant improvements in R@1 and MRR@10, and excels in intermediate ranking (R@5), outperforming models like ROGER-Ultron (TU) and LTRGR (SI). This highlights its ability to identify relevant documents early while maintaining strong intermediate performance. However, at broader recall levels (R@10), ROGER-Ultron (TU) shows a slight advantage over DDRO (TU).

\item On NQ, \ac{DDRO} (PQ) achieves the highest early precision (R@1, R@5), effectively ranking relevant documents for complex queries. The PQ-based docids capture deeper semantic relationships, leading to strong intermediate performance and competitive broader recall (R@10) against hybrid models such as ROGER-NCI (SI).
\end{enumerate*}

Overall, \ac{DDRO} consistently achieves strong performance without relying on auxiliary reward models, reinforcement learning, or dense retrieval signals. Instead, it employs \ac{SFT} followed by pairwise ranking optimization to refine early-stage precision while maintaining effectiveness across different ranking depths.

\section{Conclusion}
We introduced \ac{DDRO}, a novel approach for enhancing \ac{GenIR} systems by directly aligning docid generation with document-level relevance estimation. This alignment allows \ac{GenIR} systems to effectively learn to rank and improve accuracy. We experimented with two types of docid designs and designed a lightweight direct L2R optimization algorithm on top of SFT training.
Unlike existing RL-based methods, \ac{DDRO} simplifies optimization through pairwise ranking, eliminating the need for auxiliary reward modeling and RL fine-tuning. Experiments conducted on benchmark datasets demonstrate that \ac{DDRO} offers a lightweight optimization process while achieving high performance and demonstrating robustness. However, \ac{DDRO} has limitations that offer opportunities for further improvement. 
\begin{enumerate*}[label=(\roman*)]
\item Our evaluation has primarily focused on a subset of the MS MARCO document dataset, leaving its scalability to larger and more diverse corpora, including domain-specific and multilingual datasets, yet to be explored. Addressing this limitation will be crucial to assess \ac{DDRO}'s adaptability in broader applications.

\item Our current pairwise ranking formulation relies on binary relevance judgments (relevant vs. non-relevant), which may limit its expressiveness for queries with nuanced or graded relevance levels. Future work may explore listwise ranking objectives and graded relevance supervision to more accurately capture complex retrieval intents.

\item Potential improvements could include advanced hard negative mining to enhance relevance discrimination and multi-objective optimization to balance relevance with efficiency, fairness, and diversity. 

\item A comprehensive comparison with scalable \ac{GenIR} baselines such as RIPOR and PAG remains an avenue for future research to assess \ac{DDRO}'s scalability and integration potential in large-scale retrieval scenarios.
\end{enumerate*}


\begin{acks}
We thank our colleagues at the IRLab for their support and Hansi Zeng at UMass Amherst for helpful discussions and technical guidance. Experiments for this work were supported by the Dutch Research Council (NWO) under project EINF-9550; computations were performed on the Snellius supercomputer (SURF).
This research was (partially) supported by Ahold Delhaize, through AIRLab, by the Dutch Research Council (NWO), under project numbers 024.004.022, NWA.1389.20.\-183, and KICH3.LTP.20.006, and by the European Union's Horizon Europe program under grant agreement No 101070212. All content represents the opinion of the authors, which is not necessarily shared or endorsed by their respective employers and/or sponsors.
\end{acks}

\balance
\bibliographystyle{ACM-Reference-Format}
\bibliography{references}

\end{document}